\documentclass[journal]{IEEEtran}

\usepackage{graphicx}
\usepackage{ascmac}

\usepackage{bm}
\usepackage{amsmath}
\usepackage{amssymb}
\usepackage{amsfonts}
\usepackage{color}
\usepackage{wrapfig}
\usepackage{url}
\usepackage{cite}
\usepackage{siunitx}
\usepackage{mathtools}

\usepackage{diagbox}
\usepackage{enumerate}

\usepackage{comment}

\ifCLASSINFOpdf

\else

\fi

\makeatletter
\def\ps@IEEEtitlepagestyle{
  \def\@oddfoot{\mycopyrightnotice}
  \def\@evenfoot{}
}
\def\mycopyrightnotice{
  \begin{minipage}{\textwidth}
  \centering \scriptsize
 \copyright~2023 IEEE. Personal use of this material is permitted. Permission from IEEE must be obtained for all other uses, in any current or future media, including reprinting/republishing this material for advertising or promotional purposes, creating new collective works, for resale or redistribution to servers or lists, or reuse of any copyrighted component of this work in other works.
  \end{minipage}
}

\begin{document}

\title{A Control-theoretic Model for Bidirectional Molecular Communication Systems}

\author{Taishi~Kotsuka,~
        Yutaka~Hori,~\IEEEmembership{Member,~IEEE}
        
\thanks{This work was supported in part by JSPS KAKENHI Grant Numbers JP21H05889 and JP22J10554, and in part by JST SPRING Grant Number JPMJSP2123.}
\thanks{T. Kotsuka and Y. Hori are with the Department
of Applied Physics and Physico-Informatics, Keio University, Kanagawa 223-8522 Japan. Correspondence should be addressed to Y. Hori (email: tkotsuka@keio.jp; yhori@appi.keio.ac.jp).}

}

\maketitle

\begin{abstract}
Molecular communication (MC) enables cooperation of spatially dispersed molecular robots through the feedback control mediated by diffusing signal molecules. However, conventional analysis frameworks for the MC channels mostly consider the dynamics of unidirectional communication, lacking the effect of feedback interactions. In this paper, we propose a general control-theoretic modeling framework for bidirectional MC systems capable of capturing the dynamics of feedback control via MC in a systematic manner. The proposed framework considers not only the dynamics of
molecular diffusion but also the boundary dynamics at the
molecular robots that captures the lag due to the molecular transmission/reception process affecting the performance of the entire feedback system. Thus, methods in control theory can be applied to systematically analyze various dynamical properties of the feedback system. We perform a frequency response analysis based on the proposed framework to show a general design guideline for MC channels to transfer signal with desired control bandwidth. Finally, these results are demonstrated by showing the step-by-step design procedure of a specific MC channel satisfying a given specification.
\end{abstract}


\IEEEpeerreviewmaketitle

\section{Introduction}

In nature, bacterial cells are known to control collective behaviors of their populations via a molecular communication (MC) mediated by the diffusion of signal molecules \cite{Teresa2000,hagen2015,waters2005quorum}. Recently, this naturally existing mechanism has inspired researchers to engineer diffusion-based MC systems that enable artificially designed molecular robots to communicate with each other in a fluidic medium \cite{tatsuya2018molecular,Farsad2016,Bi2021,Soldner2020}. The diffusion-based MC has potential to enable multiple molecular robots to form a feedback system via signal molecules, and thus, it facilitates to generate cooperative behaviors, broadening the application fields of molecular robots to targeted drug delivery and remote health diagnosis \cite{Femminella2015,Gao2014,Soldner2020,Din2016}. For the streamlined design of such MC systems, fundamental properties of systems such as frequency response, stability, and robustness need to be analyzed. 
Thus, there is a need for the development of feedback control theory for bidirectional MC systems. 

\smallskip
\par
Many of the existing systems theory on MC focused on the frequency response analysis of MC channels since securing the desired control bandwidth is a critical and challenging task in the design of the diffusion-based MC due to the low-pass effect of diffusion \cite{Pierobon2010,Chude-Okonkwo2015,Huang2021}. In control engineering, control bandwidth refers to the range of frequencies within which a feedback control system can effectively operate. In other words, the MC system can respond quickly and make effective actions only to the changes within the control bandwidth. 
In \cite{Pierobon2010}, the gain characteristics and the group delay of the unidirectional MC channel was analyzed based on Green's function of diffusion equation. However, this approach is not directly applicable to the analysis of bidirectional MC channels since bidirectional MC channels can cause self-interference by receiving signal molecules transmitted by the molecular robot itself, and the frequency response characteristics of bidirectional MC channels would be different from that of conventional unidirectional MC channels. To analyze the effect of bidirectional communication, a feedback model was previously proposed from the authors' group by combining unidirectional MC channels \cite{Hara2021,kotsuka2022frequency}. Applications of this model were, however, limited due to its standing assumption that signal reception is based on the receptors on the molecular robots, which makes it hard to capture transmission/reception mechanisms by secretion and absorption of signal molecules such as quorum sensing, one of the major methodologies to implement MC.
This motivates to develop a unified system theoretic framework for the modeling, analysis and design of bidirectional molecular communication systems.

\smallskip
\par
In this paper, we propose a control-theoretic modeling framework for bidirectional MC channels with two molecular robots and demonstrate the streamlined design process of MC systems based on their frequency response analysis. The proposed framework considers not only the dynamics of molecular diffusion but also the boundary dynamics that captures the lag of signal transfer due to the molecular transmission/reception process at the interface between the communication channel and the molecular robots, which affects the performance of the entire feedback system. The overall model, thus, consists of a feedback loop of molecular robots, the boundary system, and the diffusion-based MC channel. This model enables comprehensive analysis and design of the dynamics of the feedback systems using methods in control theory. Using the proposed model, we analyze the frequency response characteristics of the bidirectional MC channel based on the transfer functions. In particular, we show a general design guideline of the MC channel and the membrane of the molecular robots so that signals with pre-specified control bandwidth is transferred while undesirable effects of the self-interference are suppressed. Finally, we perform the frequency response analysis for a specific MC channel to demonstrate these results.

\smallskip
\par
This paper is organized as follows. In the next section, we model bidirectional MC systems and introduce general control-theoretic models of the MC systems consisting of three subsystems; the diffusion system, the boundary systems, and the molecular robots. In Section \ref{sec:transfer}, we analyze the frequency response characteristics of the diffusion system to design the open loop characteristics of the MC channel based on the cut-off frequency of the diffusion system. We then provide the design guideline of the boundary systems to suppress the undesirable effect of the self-interference in Section \ref{sec:boundarydiffusion}. These results are then demonstrated in Section \ref{sec:numerical} for the design of a specific MC system. Finally, the paper is concluded in Section \ref{sec:conclusion}.

\smallskip
\noindent
\textbf{Notations:}
The $(i,j)$-th entry of a matrix $A$ is denoted by as $A_{ij}$. $\mathbb{R}$ is a set of real numbers, $\mathbb{R}^n$ is a set of $n$ dimensional vectors with real entries.

\section{Control-theoretic model of MC systems}
\label{sec:topology}

In this section, we model bidirectional MC systems consisting of the diffusion-based MC channel with two molecular robots. We then introduce the general control-theoretic MC model that expresses the feedback interaction of the diffusion region, the molecular robots, and their boundaries in the MC system.

\subsection{Governing equations}

\begin{figure}
    \centering
    \includegraphics[width=0.99\linewidth]{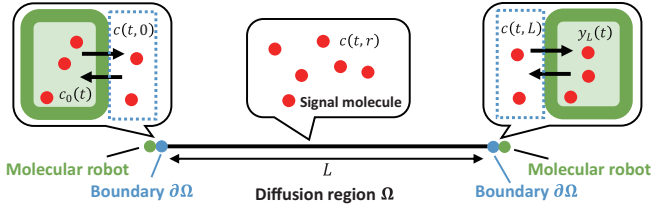}
    \caption{The molecular communication system. The system consists of a bidirectional MC channel with two molecular robots at both ends. Both molecular robots communicate by sending and receiving the same species of signal molecules to and from each other.}
    \label{fig:molecularcommun}
\end{figure}

We consider a bidirectional MC channel with two molecular robots at both ends as shown in Fig. \ref{fig:molecularcommun}. Both molecular robots communicate with each other by signal molecules diffusing in one-dimensional fluidic environment $\Omega = [0,L]$ with $L$ being the communication distance. The MC system consists of three parts: the diffusion region, the molecular robots, and the boundaries between them.

\smallskip
\par
In the diffusion region, the signal molecules diffuse in fluidic environment, which is modeled by diffusion equation with the molecular concentration of the signal molecules $c(t,r)$ at time $t$ and position $r$ as
\begin{eqnarray}
\frac{\partial c(t,r)}{\partial t} &=& \mu\frac{\partial^2 c(t,r)}{\partial r^2},\label{eq:dif}
\end{eqnarray}
where $\mu$ is the diffusion coefficient. We assume that the initial condition is
\begin{eqnarray}
c(0,r) &=& 0; \ \forall r \in \Omega,\label{eq:initial}
\end{eqnarray}

\smallskip
\par
\begin{figure}
    \centering
    \includegraphics[width=0.95\linewidth]{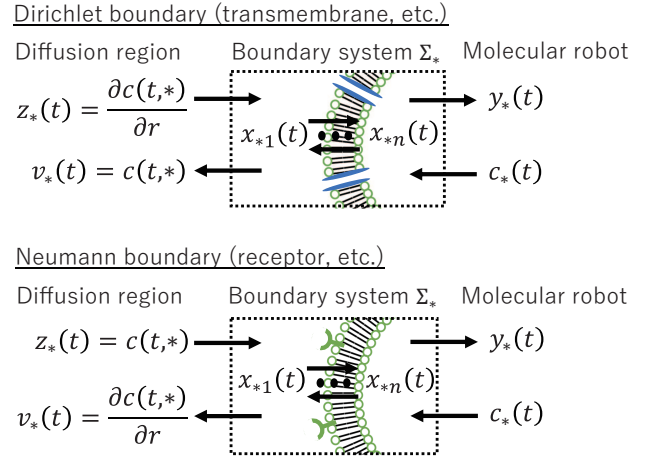}
    \caption{Dynamics at the boundary between the diffusion region and the molecular robot for different membrane signal transfer mechanisms.}
    \label{fig:boundary}
\end{figure}

The boundary conditions depend on the membrane signal transfer mechanisms at both ends since the molecular concentration at the boundaries changes dynamically based on the membrane signal transfer and the inflow/outflow of the signal molecules from/to the diffusion region. Let $\partial \Omega$ denote the boundary of the region $\Omega$ at position $*$ ($* = 0, L)$.  We define the boundary system $\Sigma_*$ that captures the dynamic input-output relation between the inside of the molecular robot at position $*$ and the diffusion region as illustrated in Fig. \ref{fig:boundary}, respectively. More specifically, the boundary system $\Sigma_*$ represents a dynamic Dirichlet boundary condition when the membrane signal transfer mechanism is the secretion and the absorption of molecules, and it represents a dynamic Neumann boundary condition in the case of ligand-receptor reception. For many practical examples, the boundary system $\Sigma_*$ can be modeled by a linear time-invariant system of the form 
\begin{equation}
\begin{split}
\frac{d\bm{x_{*}}(t)}{dt} &= A_*\bm{x_{*}}(t) + B_{*}\left[\begin{matrix}\displaystyle z_*(t)\\\displaystyle c_*(t)\end{matrix}\right], \label{eq:blayer}\\
\left[\begin{matrix}\displaystyle v_*(t)\\\displaystyle y_*(t)\end{matrix}\right] &= C_{*}\bm{x_{*}}(t) + D_*\left[\begin{matrix}\displaystyle z_*(t)\\\displaystyle c_*(t)\end{matrix}\right],
\end{split}
\end{equation}
where $A_*\in\mathbb{R}^{n\times n}$, $B_*\in\mathbb{R}^{n\times 2}$, $C_*\in\mathbb{R}^{2\times n}$, and $D_*\in\mathbb{R}^{2\times 2}$. The state $\bm{x_{*}}(t)\in\mathbb{R}^n$ is the concentration of molecules associated with the membrane signal transfer, $y_*(t)$ is the signal transmitted into the molecular robot as a result of membrane signal transfer as shown in Fig. \ref{fig:boundary}, and $c_*(t)$ is the total concentration of the signal molecules in the molecular robot. The physical correspondence of the variables $v_*(t)$ and $z_*(t)$ depends on the membrane signal transfer mechanisms as seen in Fig. \ref{fig:boundary}. Specifically, $v_*(t)$ and $z_*(t)$ are either the concentration $c(t,*)$ or the concentration gradient $\partial c(t,*)/\partial r$, which is $\partial c(t,r)/\partial r$ at position $*$ defined by 
\begin{align}
\begin{cases}
\displaystyle
v_*(t) = c(t, *),\  z_*(t) = \frac{\partial c(t,*)}{\partial r} & (\partial \Omega \ \mathrm{is\ Dirichlet})\\
\displaystyle 
v_*(t) = \frac{\partial c(t,*)}{\partial r},\  z_*(t) = c(t, *) & (\partial \Omega \ \mathrm{is\ Neumann}).
\end{cases}
\label{eq:v-z-def}
\end{align}

\par
\smallskip
There are several mechanisms to transfer signal molecules via membrane including transmembrane and ligand-receptor systems. In the following examples, we show that the dynamics of typical membrane signal transfer mechanisms can be represented by the state space model (\ref{eq:blayer}). 

\par
\smallskip
\noindent
{\bf Example 1.} We consider transmembrane systems such as quorum sensing mechanisms and ion channels \cite{Pai2009, lodish2016molecular}, which passively transport specific molecules depending on the difference of the concentration between the inside and the outside of a molecular robot. The boundary system $\Sigma_0$ can then be modeled based on the law of conservation of mass as
\begin{eqnarray}
\frac{dc_{\mathrm{out}}(t)}{dt} &=& k(c_0(t) - c_{\mathrm{out}}(t)) + \mu\frac{\partial c(t,0)}{\partial r}, \nonumber\\
y_0(t) &=& k(c_{\mathrm{out}}(t) - c_0(t)),
\label{eq:transmembrane}
\end{eqnarray}
where $c_{\mathrm{out}}(t)$ is the concentration of the signal molecules outside of the molecular robot, $k$ is the membrane transport rate. $y_0(t)$ is the flow of the signal molecules into the molecular robot. The transmembrane dynamics (\ref{eq:transmembrane}) can then be translated into the state space model (\ref{eq:blayer}) as
\begin{eqnarray}
\frac{dx_{0}(t)}{dt} &=& -k x_{0}(t) + \left[\begin{matrix}
\displaystyle \mu&k
\end{matrix} \right]\left[\begin{matrix}\displaystyle \frac{\partial c(t,0)}{\partial r}\\\displaystyle c_0(t)\end{matrix}\right]
,\nonumber\\
\left[\begin{matrix}c(t,0)\\y_0(t)\end{matrix}\right] &=& \begin{bmatrix}
1\\k
\end{bmatrix}x_{0}(t) + \begin{bmatrix}
0&0\\0&-k
\end{bmatrix} \left[\begin{matrix}\displaystyle \frac{\partial c(t,0)}{\partial r}\\\displaystyle c_0(t)\end{matrix}\right] \label{eq:mtrans}.
\end{eqnarray}

\par
\smallskip
\noindent
{\bf Example 2.} We next consider ligand-receptor systems, where the molecular robot receives signals via receptors. The signal molecules arriving at the boundary $\partial\Omega$ adsorb onto receptors on the surface of the molecular robot and then desorb. The dynamics of the boundary system $\Sigma_L$ can be represented as
\begin{eqnarray}
\frac{dc_A(t)}{dt} &=& -\mu\frac{\partial c(t,L)}{\partial r} = - k_{\mathrm{off}}c_A(t) + Rk_{\mathrm{on}}c(t,L),\label{eq:recept}\\
y_L(t) &=& k_{\mathrm{re}}c_A(t),\nonumber
\end{eqnarray}
where $c_A(t)$ is the molecular concentration adsorbed by receptors \cite{Lauffenburger1996receptors,Andrews2009,Bi2021}. The constants $k_{\mathrm{on}}$ and $k_{\mathrm{off}}$ are the adsorption/desorption rate, respectively, and $k_{\mathrm{re}}$ is the signal transduction rate into the molecular robot. The constant $R$ is the total number of receptors, assuming that it is much larger than the number of the signal molecule/receptor complexes. The output $y_L(t)$ is the signal into the molecular robot, which is proportional to $c_A(t)$. The state space model (\ref{eq:blayer}) can be expressed based on the ligand-receptor dynamics (\ref{eq:recept}) as
\begin{eqnarray}
\frac{dx_L(t)}{dt} &=&  - k_{\mathrm{off}}x_L(t) + [\begin{matrix}k_{\mathrm{on}}&0\end{matrix}]\left[\begin{matrix}c(t,L)\\c_L(t)\end{matrix}\right],
\label{eq:receptor}\\
\left[\begin{matrix}\displaystyle\frac{\partial c(t,L)}{\partial r}\\y_L(t)\end{matrix}\right] &=& \left[\begin{matrix}\displaystyle-\frac{k_{\mathrm{off}}}{\mu}\\k_{\mathrm{re}}\end{matrix}\right]x_L(t) + \left[\begin{matrix}\displaystyle\frac{k_{\mathrm{on}}}{\mu}&0\\0&0\end{matrix}\right]\left[\begin{matrix}c(t,L)\\c_L(t)\end{matrix}\right],\nonumber
\end{eqnarray}
where $x_L(t)=c_A(t)$.

\subsection{Control-theoretic model in frequency domain}

Next, we introduce a unified control-theoretic model of the MC system that represents the feedback interaction of the three components, {\it i.e.,} molecular robots, diffusion system, and the boundary systems, of the MC systems in general.

\par
\begin{figure}
    \centering
    \includegraphics[width=0.99\linewidth]{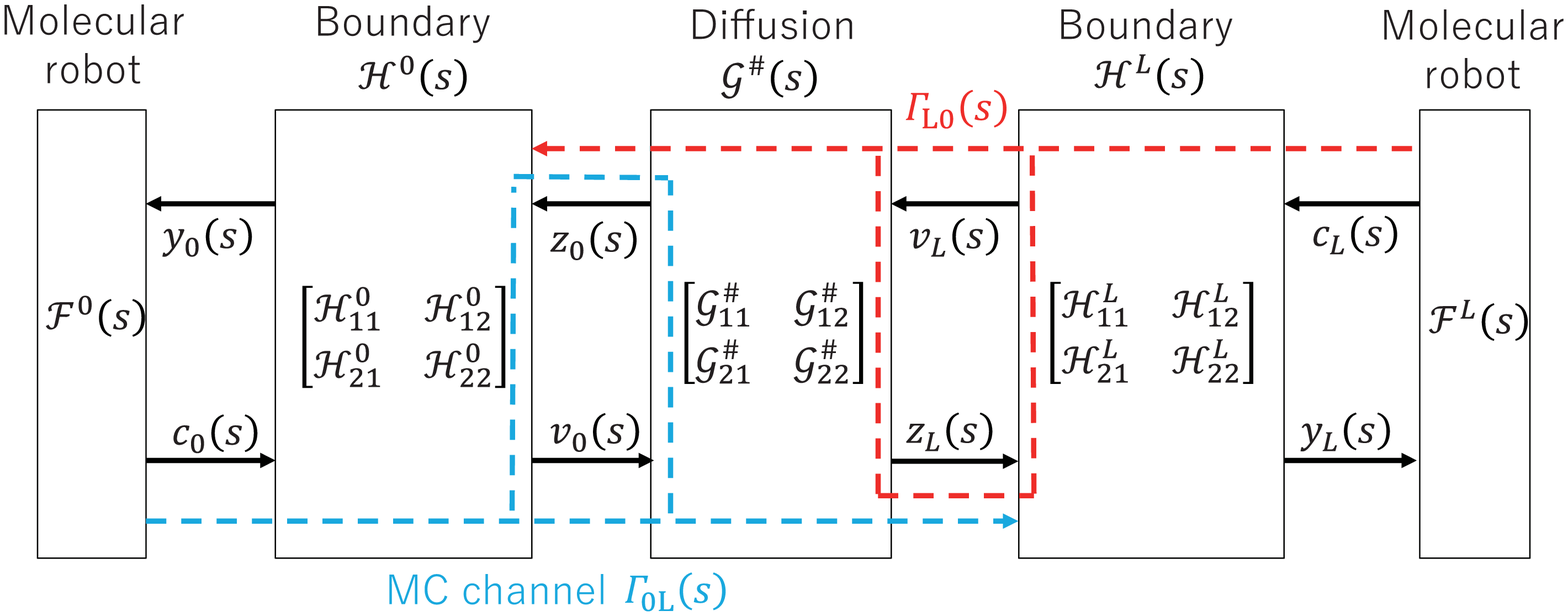}
    \caption{Control-theoretic model of the MC system. The model represents the feedback interaction of the three components, the diffusion system $\mathcal{G}^{\sharp}(s)$, the boundary systems $\mathcal{H}^*(s)$, and the molecular robots $\mathcal{F}^*(s)$.}
    \label{fig:blockabstract}
\end{figure}

\smallskip
Let $\mathcal{G}^{\sharp}(s)$ denote the multi-input multi-output transfer function of the diffusion system, where the two input and the output is defined as $[v_0(s), v_L(s)]^T$ and $[z_0(s), z_L(s)]^T$, respectively. The subscript $\sharp=\mathrm{dn}, \mathrm{dd}, \mathrm{nd}, \mathrm{nn}$ represents the different combinations of the boundary mechanisms at both ends of the region $\Omega$, where $\mathrm{d}$ is Dirichlet boundary and $\mathrm{n}$ is Neumann boundary. We also define $\mathcal{H}^*(s)$ and $\mathcal{F}^*(s)$ as the transfer functions of the boundary systems and its associated molecular robots at position $* (* = 0, L)$, respectively. Then, the MC system can be systematically modeled as the feedback system shown in Fig. \ref{fig:blockabstract} with the three types of subsystems: the diffusion system $\mathcal{G}^{\sharp}(s)$, the boundary systems $\mathcal{H}^*(s)$, and the reaction system in the molecular robots $\mathcal{F}^*(s)$. The feedback interaction of the subsystems is formally written as

\begin{eqnarray}
\left[\begin{matrix} z_0(s)\\
z_L(s)\end{matrix}\right] &=& \mathcal{G}^{\sharp}(s)\left[\begin{matrix} v_0(s)\\
v_L(s)\end{matrix}\right],\label{eq:difsys}\\
\left[\begin{matrix} v_*(s)\\
y_*(s)\end{matrix}\right] &=& \mathcal{H}^*(s)\left[\begin{matrix} z_*(s)\\
c_*(s)\end{matrix}\right],\label{eq:mtransys}\\
c_*(s) &=& \mathcal{F}^*(s)c(s,*),\label{eq:cellsys}
\end{eqnarray}
where $c(s,*)$, $v_*(s)$, and $z_*(s)$ are defined as the Laplace transform of the molecular concentration $c(t,*)$, the input $v_*(t)$, and the output $z_*(t)$, respectively. 
The transfer function of the boundary system $\mathcal{H}^*(s)$ is obtained from the state space representation (\ref{eq:blayer}) as $\mathcal{H}^*(s) = C_*(sI-A_*)^{-1}B_*+D_*$\ ($*=0,L$). 
The transfer function of the diffusion system $\mathcal{G}^{\sharp}(s)$ depends on the boundary mechanisms, which determines the input $v_*(s)$ and the output $z_*(s)$ of the diffusion system $\mathcal{G}^{\sharp}(s)$ as illustrated in Fig. \ref{fig:boundary}. Specific forms of the transfer function of the diffusion system $\mathcal{G}^{\sharp}(s)$ are introduced in Section \ref{sec:transfer}. 

\par
\smallskip
The control-theoretic model (\ref{eq:difsys})--(\ref{eq:cellsys}) allows us to analyze and design each subsystem that ensures the stability, robustness and performance of the bidirectional MC systems using methods in control theory. 
In particular, the proposed model explicitly considers the interaction between the diffusion system $\mathcal{G}^{\sharp}(s)$ and the boundary system $\mathcal{H}^*(s)$, which is often overlooked despite a large impact on the overall dynamics as illustrated in Section \ref{sec:numerical}. 
Thus, we can, for example, design the frequency response characteristics of the closed-loop MC channels $\Gamma_{
\mathrm{0L}}(s)$ and $\Gamma_{
\mathrm{0L}}(s)$ in Fig. \ref{fig:blockabstract}, which are transfer functions from $c_0(s)$ to $z_L(s)$ and from $c_L(s)$ to $z_0(s)$, respectively, through the design of the boundary mechanisms at the surface of the molecular robots. 
In the following section, we first derive specific forms of the transfer functions $\mathcal{G}^{\sharp}(s)$ and analyze the effect of the feedback interaction between the diffusion system $\mathcal{G}^{\sharp}(s)$ and the boundary system $\mathcal{H}^*(s)$ on the closed-loop frequency response. This analysis then gives an insight into the design of the boundary system $\mathcal{H}^*(s)$ that attenuates undesirable effect of the boundary-diffusion interaction. 

\smallskip
\noindent

{\bf Remark 1.} 
The control-theoretic model in Fig. \ref{fig:blockabstract} also captures the effect of the cross-talk due to the use of the same molecular species between the two molecular robots.
Here, the cross-talk refers to the interference of the communication from the molecular robot at $r=0$ to the robot at $r=L$ by signals transmitted from the robot at $r=L$. 
In Fig. \ref{fig:blockabstract}, the signal $y_L(s)$ is expressed as $y_L(s) = \mathcal{M}_{0L}(s)c_0(s) + \mathcal{M}_{LL}(s)c_L(s)$, where $ \mathcal{M}_{0L}$ and $\mathcal{M}_{LL}(s)$ are the transfer functions from the inputs $c_0(s)$ and $c_L(s)$ to the output $y_L(s)$, respectively. The first term represents the transmission characteristics of the signal from the molecular robot at $r=0$ to the robot at $r=L$ while the second term $\mathcal{M}_{22}(s)c_L(s)$ represents the transmission characteristics of an undesired signal corresponding to the effect of the cross-talk, {\it i.e.}, the effect of the signal transmitted from the molecular robot at $r=L$.

\section{Frequency response analysis of the diffusion system}
\label{sec:transfer}

In this section, we show the derivation of the transfer function of the diffusion system $\mathcal{G}^{\sharp}(s)$ to analyze its frequency response characteristics. We then show the parameter dependency of the bandwidth of signals that can be transferred by passive diffusion based on the transfer function. 

\subsection{Transfer function of the diffusion system}
\label{sec:trasmodule}

Recall that the entries of the transfer function matrix $\mathcal{G}^{\sharp}(s)$ consist of the transfer functions from $v_*(s)$ to $z_*(s)$ $(* = 0, L)$. The $v_*(s)$ and $z_*(s)$ are either the concentration or concentration gradient of the signal molecule at the boundary $\partial\Omega$ as defined in (\ref{eq:v-z-def}).
Here, we derive a general form of these transfer functions and provide its system-theoretic interpretation. The following theorem shows a specific form of the transfer function from the input $v_*(s)$ to the output $z_\ell(s)$, where $z_\ell(s)$ is the Laplace transform of the concentration $c(t, \ell)$ or its  gradient $\partial c(t, r)/\partial r$ at position $\ell \in \Omega$.

\par
\medskip
\noindent
{\bf Theorem 1.}
Consider the diffusion equation (\ref{eq:dif}) with the initial condition (\ref{eq:initial}) and the boundary conditions (\ref{eq:v-z-def}). The transfer function from $v_*(s)$ to $z_\ell(s)$ is given by $\mathcal{Z}_\ell G_{r, *}(s)$ $(* = 0, L)$, \textit{i.e.,}
\begin{align}
z_\ell(s) = \mathcal{Z}_\ell G_{r, *}(s) v_*(s), \label{eq:zr}
\end{align}
where 
\begin{align}
&G_{r, *}(s) = \frac{g(s,L-d)Kg(s,L)+g(s,d)}{1 + Kg(s,L)g(s,L)}\label{eq:difsyssr}
\end{align}
with
\begin{align}
&g(s,r) = e^{-\frac{r}{\sqrt{\mu}}\sqrt{s}}, \\
&K = 
\begin{cases}
-1 & (\text{$\partial \Omega$ are both Dirichlet or both Neumann})\\
1 & \text{otherwise},
\end{cases} \\
&d = |r - *|.
\end{align}
The operator $\mathcal{Z}_{\ell}$ on a function $f$ is defined by 
\begin{equation}
    \mathcal{Z}_\ell f = \begin{cases}
\displaystyle\frac{\partial f}{\partial r}\biggr|_{r=\ell} & \left(v_*(s)=c(s,*),\,z_\ell(s)=\displaystyle\frac{\partial c(s,r)}{\partial r}\biggr|_{r=\ell}\right) \\
\displaystyle\int^\ell_0 {f dr} & \left(v_*(s)=\displaystyle\frac{\partial c(s,r)}{\partial r}\biggr|_{r=*},\,z_\ell(s)=c(s,\ell)\right)\\
f\biggr|_{r=\ell} & \mathrm{otherwise},
\end{cases}
\end{equation}

\par
\medskip

The proof of this theorem is shown in Appendix \ref{sec:dtf}. Fig. \ref{fig:dblock} illustrates the block diagram of the transfer function $\mathcal{Z}_{\ell} G_{r, *}(s)$ in (\ref{eq:difsyssr}). This figure illustrates that the transfer function from $v_*(s)$ to $z_\ell(s)$ can be decomposed into the feedback connection of the simple transfer function $g(s,r)$, where  $g(s,r)$ is the transfer function of passive diffusion for distance $r$ with the boundary condition $\lim_{r \rightarrow \infty} c(t, r) = 0$ in one-dimensional space $\Omega=[0,\infty)$. The coefficient $K$ is the reflection coefficient of the boundary, which is $-1$ when the boundary $\partial\Omega$ is fixed end relative to the input $v_*(s)$ and 1 when the boundary is free end relative to the input $v_*(s)$.

With Theorem 1, the transfer function of the diffusion system $\mathcal{G}^{\sharp}(s)$ can be systematically obtained by substituting $\ell=0$ and $\ell=L$ into $\mathcal{Z}_{\ell}G_r,*(s)$ in Eq. (\ref{eq:zr}), \textit{i.e.,}
\begin{align}
\mathcal{G}^{\sharp}(s)=
\begin{bmatrix}
\mathcal{Z}_0 G_{r,0}(s) & \mathcal{Z}_0 G_{r,L}(s) \\
\mathcal{Z}_L G_{r,0}(s) & \mathcal{Z}_L G_{r,L}(s)
\end{bmatrix}.
\end{align}

\begin{figure}
    \centering
    \includegraphics[width=0.99\linewidth]{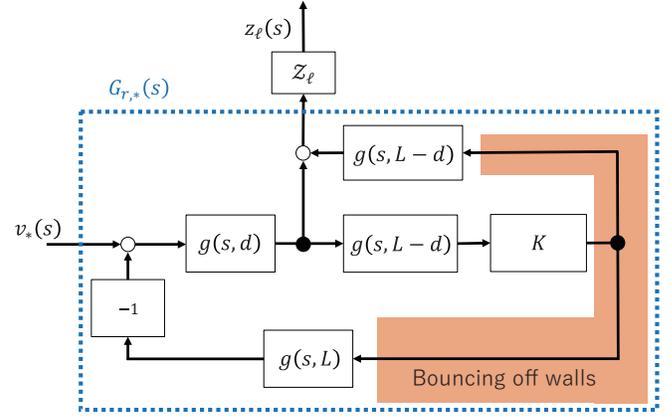}
    \caption{The block diagram of the diffusion process.}
    \label{fig:dblock}
\end{figure}

\smallskip
\par

\smallskip
\par
\noindent
{\bf Example 3.} 
Consider the diffusion system (\ref{eq:difsys}) with the boundary at $r = 0$ being the membrane transport and $r=L$ being the ligand-receptor mechanism. 
The input and the output of the diffusion system is then defined as 
\begin{eqnarray}
\left[\begin{matrix}v_0(s)\\v_L(s)\end{matrix}\right] &=& \left[\begin{matrix}c(s,0)\\ \displaystyle\frac{\partial c(t,L)}{\partial r}\end{matrix}\right],  \notag \\
\left[\begin{matrix}z_0(s)\\z_L(s)\end{matrix}\right] &=& \left[\begin{matrix}\displaystyle\frac{\partial c(t,0)}{\partial r}\\c(t,L)\end{matrix}\right],
\end{eqnarray}
respectively. The reflection coefficient is $K=1$ since the boundary $\partial\Omega$ at $r=0$ is Dirichlet boundary and at $r=L$ is Neumann boundary. Using Theorem 1 and substituting $\ell=0$ and $\ell=L$, the transfer function $\mathcal{G}^{\mathrm{dn}}(s)$ is obtained as

\begin{equation}
\mathcal{G}^{\mathrm{dn}}(s) = \begin{bmatrix}
\displaystyle-\sqrt{\frac{s}{\mu}} \tanh{\left(\frac{L}{\sqrt{\mu}}\sqrt{s}\right)}&\displaystyle\frac{1}{\cosh{\left(\frac{L}{\sqrt{\mu}}\sqrt{s}\right)}}\\\displaystyle\frac{1}{\cosh{\left(\frac{L}{\sqrt{\mu}}\sqrt{s}\right)}}&\displaystyle\sqrt{\frac{\mu}{s}} \tanh{\left(\frac{L}{\sqrt{\mu}}\sqrt{s}\right)}
\end{bmatrix}.
\label{eq:difsysdnindex}
\end{equation}
Four types of transfer functions $\mathcal{G}^{\sharp}(s)$ are summarized in Table \ref{tab:transform}.

\begin{table*}
\begin{center}
\caption{Transformation of the diffusion system}
\label{tab:transform}
\begin{tabular}{|c|cc|}\hline
 From\textbackslash To&Dirichlet boundary&Neumann boundary\\ \hline
 & & \\
\begin{tabular}{l}Dirichlet\\ boundary\end{tabular}& \begin{tabular}{l}$\mathcal{G}^{\mathrm{dd}}(s)=$ \\[1.5ex]
$\begin{bmatrix}
\displaystyle-\sqrt{\frac{s}{\mu}} \frac{1}{\tanh{\left(\frac{L}{\sqrt{\mu}}\sqrt{s}\right)}}&\displaystyle\sqrt{\frac{s}{\mu}}\frac{1}{\sinh{\left(\frac{L}{\sqrt{\mu}}\sqrt{s}\right)}}\\\displaystyle\textcolor{red}{-}\sqrt{\frac{s}{\mu}}\frac{1}{\sinh{\left(\frac{L}{\sqrt{\mu}}\sqrt{s}\right)}}&\displaystyle\sqrt{\frac{s}{\mu}} \frac{1}{\tanh{\left(\frac{L}{\sqrt{\mu}}\sqrt{s}\right)}}
\end{bmatrix}$\end{tabular} & \begin{tabular}{l}$\mathcal{G}^{\mathrm{dn}}(s)=$ \\[1.5ex]
$\begin{bmatrix}
\displaystyle-\sqrt{\frac{s}{\mu}} \tanh{\left(\frac{L}{\sqrt{\mu}}\sqrt{s}\right)}&\displaystyle\frac{1}{\cosh{\left(\frac{L}{\sqrt{\mu}}\sqrt{s}\right)}}\\\displaystyle\frac{1}{\cosh{\left(\frac{L}{\sqrt{\mu}}\sqrt{s}\right)}}&\displaystyle\sqrt{\frac{\mu}{s}} \tanh{\left(\frac{L}{\sqrt{\mu}}\sqrt{s}\right)}
\end{bmatrix}$\end{tabular} \\ 
 & & \\
\begin{tabular}{l}Neumann\\ boundary\end{tabular}& \begin{tabular}{l}$\mathcal{G}^{\mathrm{nd}}(s)=$ \\[1.5ex]
$\begin{bmatrix}
\displaystyle\sqrt{\frac{\mu}{s}} \tanh{\left(\frac{L}{\sqrt{\mu}}\sqrt{s}\right)}&\displaystyle\frac{1}{\cosh{\left(\frac{L}{\sqrt{\mu}}\sqrt{s}\right)}}\\\displaystyle\frac{1}{\cosh{\left(\frac{L}{\sqrt{\mu}}\sqrt{s}\right)}}&\displaystyle-\sqrt{\frac{s}{\mu}} \tanh{\left(\frac{L}{\sqrt{\mu}}\sqrt{s}\right)}
\end{bmatrix}$\end{tabular} & \begin{tabular}{l}$\mathcal{G}^{\mathrm{nn}}(s)=$ \\[1.5ex]
$\begin{bmatrix}
\displaystyle-\sqrt{\frac{\mu}{s}} \frac{1}{\tanh{\left(\frac{L}{\sqrt{\mu}}\sqrt{s}\right)}}&\displaystyle\sqrt{\frac{\mu}{s}}\frac{1}{\sinh{\left(\frac{L}{\sqrt{\mu}}\sqrt{s}\right)}}\\\displaystyle\textcolor{red}{-}\sqrt{\frac{\mu}{s}}\frac{1}{\sinh{\left(\frac{L}{\sqrt{\mu}}\sqrt{s}\right)}}&\displaystyle\sqrt{\frac{\mu}{s}} \frac{1}{\tanh{\left(\frac{L}{\sqrt{\mu}}\sqrt{s}\right)}}
\end{bmatrix}$
\end{tabular} \\
 & & \\ \hline
\end{tabular}
\\[1.2ex]
\end{center}
\end{table*}

\subsection{Frequency response characteristics of the diffusion system}

The (1,2) and (2,1) entries of $\mathcal{G}^{\sharp}(s)$, $\mathcal{G}^{\sharp}_{12}(s)$ and $\mathcal{G}^{\sharp}_{21}$, represent the transfer functions from the boundary system of a molecular robot to another. These transfer functions, thus, determine the open loop transfer functions of the diffusion system. 
On the other hand, the (1,1) and (2,2) entries, $\mathcal{G}^{\sharp}_{11}(s)$ and $\mathcal{G}^{\sharp}_{22}(s)$, represent the self-feedback effect due to the reflection of the signal at the terminal, which forms a feedback between the boundary system $\mathcal{H}^*(s)$ and the diffusion system $\mathcal{G}^{\sharp}(s)$ and complicates the design of the entire MC system. 
Thus, we here consider a design scenario, where we translate specifications of the MC system into the open loop transfer functions.

To this end, we first analyze the frequency response characteristics of the (1,2) and (2,1) entries of the diffusion systems $\mathcal{G}^{\sharp}(s)$ based on Eq. (\ref{eq:difsysdnindex}) and Table \ref{tab:transform}. Specifically, we analyze the cut-off frequency $\omega_D$ of the transfer functions $\mathcal{G}^{\sharp}_{21}(s)$ and $\mathcal{G}^{\sharp}_{12}(s)$ to show the relation between the bandwidth of signals that can be transferred by diffusion and the parameters such as the communication distance $L$ and the diffusion coefficient $\mu$. In what follows, we first show a proposition that the gain of the transfer functions $\mathcal{G}^{\sharp}_{21}(j\omega)$ and $\mathcal{G}^{\sharp}_{12}(j\omega)$ monotonically decreases. This proposition then facilitates the analysis of the relation between the cut-off frequency and the parameters of the communication channel. 

\par
\smallskip
\noindent
{\bf Proposition 1.} Consider the diffusion systems $\mathcal{G}^{\mathrm{dd}}(s)$, $\mathcal{G}^{\mathrm{dn}}(s)$, $\mathcal{G}^{\mathrm{nd}}(s)$, and $\mathcal{G}^{\mathrm{nn}}(s)$ defined in Table \ref{tab:transform}. The gain of the transfer functions at (1,2) and (2,1) entries of these transfer functions monotonically decreases for $\omega\geq0$, \textit{i.e.,}
\begin{align}
\frac{d |\mathcal{G}^\sharp_{12}(j\omega)|}{d \omega} < 0,\ 
\frac{d |\mathcal{G}^\sharp_{21}(j\omega)|}{d \omega} < 0\ ; \forall \omega \in [0, \infty), 
\label{eq:gain}
\end{align}
where $\mathcal{G}^\sharp_{ij}(j\omega)$ represents the $(i,j)$-th entry of $\mathcal{G}^\sharp(s)$ with $\sharp = \text{dd}, \text{dn}, \text{nd}, \text{nn}$.

\smallskip
\par
The proof of Proposition 1 is in Appendix B. 
This proposition implies that the cut-off frequency $\omega_D$ is uniquely determined, if it exists, because of the monotone property of the gain (\ref{eq:gain}). Moreover, $L$ and $\mu$ can be eliminated from the gain of $\mathcal{G}^\sharp_{12}(j\omega)$ and $\mathcal{G}^\sharp_{21}(j\omega)$ by normalizing the frequency as $\hat{\omega}:=L^2\omega/\mu$. Thus, the cut-off frequency can be calculated by binary search over the normalized frequency $\hat{\omega} \in [0, \infty)$. 

The cut-off frequency and the steady gain for $\mathcal{G}^\sharp_{12}(j\omega)$ and $\mathcal{G}^\sharp_{21}(j\omega)$ are summarized in Table \ref{tab:cutoff}. In Table \ref{tab:cutoff}, the cut-off frequency is defined as the smallest frequency, where the gain reaches $-6$ dB, which means that the amplitude of the molecular concentration is reduced by half. The cut-off frequencies for $\mathcal{G}^{\mathrm{dd}}_{21}(j\omega)$ and $\mathcal{G}^{\mathrm{nn}}_{21}(j\omega)$ could not be obtained since $L$ and $\mu$ are not eliminated by normalization of the frequency. However, for $\mathcal{G}^{\mathrm{dd}}_{21}(j\omega)$, the frequency at which the gain decreases by 6 dB from the steady gain can be found as shown in parenthesis in Table \ref{tab:cutoff}.
These results illustrate the \textit{open loop} characteristics of the diffusion system that we can use for designing desired MC systems with pre-specified control bandwidth.
Specifically, the admissible communication distance $L$ and diffusion coefficient $\mu$ can be specified based on Table \ref{tab:cutoff} to enable signal transfer with the desired bandwidth.

\begin{table}
\begin{center}
\caption{Frequency response characteristics of open loop diffusion systems}
\label{tab:cutoff}
\begin{tabular}{ccccc}\hline\hline
&From&To&Steady gain&Cut-off freq.\\ \hline\hline
\\[0.1ex]
$\mathcal{G}^{\mathrm{dn}}_{21}$&Dirichlet&Neumann&0 dB&$\displaystyle 4.14\frac{\mu}{L^2}$\\[2ex] 
$\mathcal{G}^{\mathrm{dd}}_{21}$&Dirichlet&Dirichlet&$\displaystyle20\log\left|\frac{1}{L}\right|$ dB&$\left(\displaystyle15.0\frac{\mu}{L^2}\right)$ \\[2ex] 
$\mathcal{G}^{\mathrm{nd}}_{21}$&Neumann&Dirichlet&0 dB&$\displaystyle 4.14\frac{\mu}{L^2}$\\[2ex] 
$\mathcal{G}^{\mathrm{nn}}_{21}$&Neumann&Neumann&$\infty$&\diagbox[height=\line]{\ }{\ } \\[1.5ex]\hline\hline
\end{tabular}
\end{center}
The value in parentheses indicates the frequency at which the gain decreases by 6 dB from the steady gain.
\end{table}

\section{Effect of self-interference}
\label{sec:boundarydiffusion}

\par
\smallskip
In this section, we analyze the frequency response characteristics of the feedback interaction between the diffusion system $\mathcal{G}^{\sharp}(s)$ and the boundary systems $\mathcal{H}^*(s)$, which is not considered in the analysis for the open loop system. We then show the design guideline of the boundary system $\mathcal{H}^*(s)$ for suppressing undesirable effect of the feedback interaction. 

\par
\smallskip
 
We consider the transfer functions of the MC channels $\Gamma_{\mathrm{0L}}(s)$ and $\Gamma_{\mathrm{L0}}(s)$, which are specifically defined as 
\begin{eqnarray}
    \Gamma_{\mathrm{0L}}(s) = \mathcal{G}^{\sharp}_{21}(s)S^0(s)\mathcal{H}^0_{12}(s), \label{eq:gam-0L}\\
    \Gamma_{\mathrm{L0}}(s) = \mathcal{G}^{\sharp}_{12}(s)S^L(s)\mathcal{H}^L_{12}(s), \label{eq:gam-L0}
\end{eqnarray}
where $S^0(s)$ and $S^L(s)$ are the self-interference systems defined by 
\begin{eqnarray}
S^0(s) &=& \frac{1}{1-\mathcal{H}^0_{11}(s)\mathcal{G}^{\sharp}_{11}(s)},\label{eq:boundarydiffusion}\\
S^L(s) &=& \frac{1}{1-\mathcal{H}^L_{11}(s)\mathcal{G}^{\sharp}_{22}(s)}.\label{eq:boundarydiffusion2}
\end{eqnarray}
The self-interference system $S^*(s)$ represents the self-feedback effect of the transmitted signal on the boundary system of the sender itself. In other words, $S^*(s)$ accounts for a feedback interaction between the diffusion system $\mathcal{G}^{\sharp}(s)$ and the boundary systems $\mathcal{H}^*(s)$. Physically, this corresponds to the situation where signal molecules secreted from a robot accumulate around the boundary of the diffusion region and make it difficult for the robot to further secrete the signal molecules due to less concentration difference between both ends of the membrane. This phenomenon is called self-interference \cite{Ahmadzadeh2015}.
As shown in Eqs. (\ref{eq:gam-0L}) and (\ref{eq:gam-L0}), the self-interference system introduces changes in the frequency characteristics of the open loop transfer functions $\mathcal{H}_{12}^{*}(s) \mathcal{G}_{21}^\sharp(s)$, and thus, the transmitted signal is potentially attenuated due to the self-interference. 

Here, we consider a design strategy that makes $|S^*(j\omega)|\approx 1$ and shapes the gain characteristics of the closed-loop transfer functions $\Gamma_{\mathrm{0L}}(s)$ and $\Gamma_{\mathrm{L0}}(s)$ based on the open loop transfer functions $\mathcal{H}_{12}^{*}(s) \mathcal{G}_{21}^\sharp(s)$. For this purpose, we analyze the frequency response characteristics of $S^*(s)$ and obtain the design criteria of the boundary system $\mathcal{H}^*(s)$ for suppressing the effect of the self-interference system. Without loss of generality, we consider the self-interference system at position $r = 0$, \textit{i.e.,} $S^0(s)$. It should be noted that the analysis result for $S^0(s)$ applies to $S^L(s)$ in the same way. To make $|S^0(j\omega)|\approx 1$, the gain of $\mathcal{H}^*_{11}(j\omega)\mathcal{G}^{\sharp}_{11}(j\omega)$ must be kept small according to Eq. (\ref{eq:boundarydiffusion}). For this purpose, the boundary systems $\mathcal{H}^*_{11}(s)$ should be designed according to the gain of $\mathcal{G}^{\sharp}_{11}(j\omega)$ to make $\mathcal{H}^*_{11}(j\omega)\mathcal{G}^{\sharp}_{11}(j\omega)$ small.

\par
\begin{figure}
    \centering
    \includegraphics[width=0.95\linewidth]{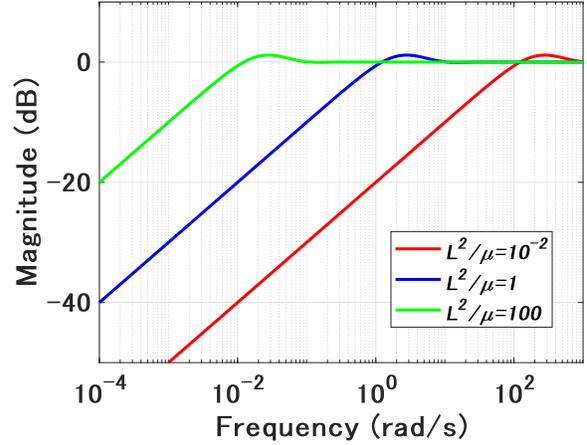}
    \caption{The gain diagram of the transfer function $\tanh{(L\sqrt{j\omega}/\sqrt{\mu})}$.}
    \label{fig:sinhcosh}
\end{figure}

\smallskip
The four types of transfer functions $\mathcal{G}^{\sharp}_{11}(s)$ shown in Table \ref{tab:transform} consist of the combinations of $\sqrt{s}/\sqrt{\mu}$ and $\tanh{(L\sqrt{s}/\sqrt{\mu})}$. 
Unlike typical transfer functions in control engineering, these functions depend on the square root of the frequency variable, \textit{$\sqrt{s}$}. The gain of $\sqrt{j\omega}/\sqrt{\mu}$ increases monotonically at 10 dB/dec for $\omega \in [0, \infty)$. The gain of $\tanh{(L\sqrt{j\omega}/\sqrt{\mu})}$, on the other hand, is approximated by 
\begin{eqnarray}
\left|\tanh{\left(\frac{L}{\sqrt{\mu}}\sqrt{j\omega}\right)}\right|\approx
\left\{
\begin{array}{ll}
1 & (|\omega| > \frac{\mu}{L^2}) \\
\frac{L}{\sqrt{\mu}}\sqrt{|\omega|} & (|\omega| \leq \frac{\mu}{L^2})
\end{array}
\right.\label{eq:tangain}.
\end{eqnarray}
This approximation is obtained by Taylor expansion at ${\omega}=0$ and 
\begin{eqnarray}
\lim_{\omega\rightarrow\infty}\left|\tanh{\left(\frac{L}{\sqrt{\mu}}\sqrt{j\omega}\right)}\right|=1.
\end{eqnarray}
 Eq. (\ref{eq:tangain}) shows that the gain of $\tanh{(L\sqrt{j\omega}/\sqrt{\mu})}$ increases at 10 dB/dec for the frequency bandwidth $[0,\mu/L^2]$, where $\mu/L^2$ is the approximate cut-off frequency for the high pass filter. The gain plot of $\tanh{(L\sqrt{j\omega}/\sqrt{\mu})}$ is shown in Fig. \ref{fig:sinhcosh}, which agrees with the approximation in Eq. (\ref{eq:tangain}). Therefore, to attenuate undesirable effect of the self-interference, the boundary system $\mathcal{H}^0_{11}(s)$ should be designed such that $|\mathcal{H}^*_{11}(j\omega)\mathcal{G}^{\sharp}_{11}(j\omega)|\ll 1$ for the control bandwidth based on the cut-off frequency of $\tanh{(L\sqrt{s}/\sqrt{\mu})}$ and the gain of $\sqrt{j\omega}/\sqrt{\mu}$. 

\par
\smallskip
\noindent

\section{Numerical example}
\label{sec:numerical}

In this section, we show a design procedure of a specific MC channel with transmembrane signal transfer in the boundary system $\Sigma_0$ and receptor-mediated membrane signal transfer in the boundary system $\Sigma_L$ capable of transferring signals with a desired control bandwidth. To this end, we first explore the parameter range of each subsystem in $\Gamma_{0L}(s)$, {\it i.e., the transfer function from $c_0(s)$ to $z_L(s)$,} that transfers signals within the desired bandwidth based on the open loop transfer function. We then show the parameter conditions to suppress the influence of the self-interference system $S^*(s)$. Finally, we expand the analysis of the MC channel $\Gamma_{0L}(s)$ to the entire MC channel $\mathcal{M}_{0L}(s)$, {\it i.e.}, the transfer function from $c_0(s)$ to $y_L(s)$, by considering the influence of the ligand-receptor system.

\subsection{Design of the MC channel $\Gamma_{0L}(s)$ for suppressing self-interference}
\label{sec:num-design}

\par
\begin{figure}
    \centering
    \includegraphics[width=0.95\linewidth]{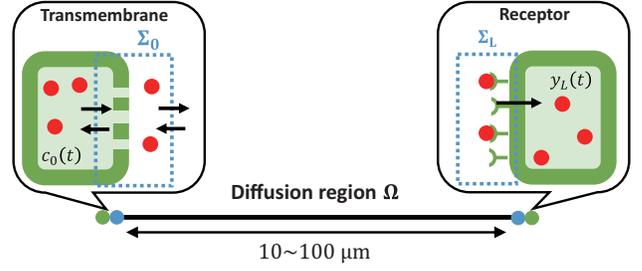}
    \caption{The MC channels with the molecular robots of transmembrane type and receptor type.}
    \label{fig:situation}
\end{figure}

\smallskip
Consider an MC system that has transmembrane signal transfer in the boundary system $\Sigma_0$ and receptor-mediated membrane signal transfer in the boundary system $\Sigma_L$ as seen in Fig. \ref{fig:situation}. The signal molecules produced in the molecular robot on the left are emitted to or absorbed from the boundary of the diffusion region $\partial \Omega$. Thus, the dynamics of the boundary system $\Sigma_0$ can be modeled by Eq. (\ref{eq:mtrans}). In the boundary system $\Sigma_L$, the signal molecules at the boundary $\partial\Omega$ are sensed by the receptors as the signal intensity that transfers into the molecular robot on the right. Here, the objective of the design problem is to explore the conditions for the membrane of the molecular robot on the left and the signal molecules that satisfy the following specifications: 
\begin{itemize}
\item[(a)] The bandwidth for control is $\varpi := [0, 10^{-2}] \mathrm{rad/s}$
\item [(b)] The communication distance $L$ is limited within $10 \,\si{\micro m} \le L \le 100\,\si{\micro m}$
\item [(c)] The thickness of the small region at the boundary $\partial \Omega$ is $\Delta r=1\,\si{\micro m}$.
\end{itemize}
To be more specific, our goal is to design the boundary system $\Sigma_0$ and the diffusion coefficient $\mu$ satisfying these conditions (a)--(c).

\par
\smallskip
To this end, we first compute the transfer function of each subsystem in the MC channel $\Gamma_{\mathrm{0L}}(s)$. The boundary system $\mathcal{H}^0(s)$ is obtained as
\begin{eqnarray}
\mathcal{H}^0(s) = \left[\begin{matrix}\displaystyle\frac{\mu}{s+k}&\displaystyle\frac{k}{s+k}\\[1.5ex]\displaystyle\frac{\mu k}{s+k}&\displaystyle-\frac{ks}{s+k}\end{matrix}\right].
\label{eq:unitrans}
\end{eqnarray}
based on Eq. (\ref{eq:mtrans}). Next, the transfer function matrix of the diffusion system is obtained as  $\mathcal{G}^{\sharp}(s)=\mathcal{G}^{\mathrm{dn}}(s)$ shown in Table \ref{tab:transform} since the two inputs of the diffusion systems from the boundary systems $\Sigma_0$ and $\Sigma_L$ are defined by $v_0(s)=c(s,0)$ and $v_L(s)=\partial c(s,L)/\partial r$, respectively (see Fig. \ref{fig:boundary}). 
Based on these transfer functions, we seek for the conditions of the membrane transport rate $k$ and the diffusion coefficient $\mu$ to achieve the design objective. The diffusion coefficient $\mu$ can be adjusted by changing the type of signal molecule \cite{Soldner2020}, and the membrane transport rate $k$ could possibly be modified by changing the number of pores on the membrane formed by proteins. For example, in \cite{Dupin2019}, protein pores such as $\alpha$-haemolysin, which inserts itself into the membrane and forms pores that spans lipid bilayers, were synthesized in the molecular robot (vesicle), leading to the increase of the membrane transport rate of small molecules outside the vesicle.

\par
\smallskip
We use the spirit of the loop shaping approach in control engineering \cite{feyel2013} to design the closed loop characteristics of the MC channel $\Gamma_{\mathrm{0L}}(s)$ using the open loop transfer function
 $\mathcal{G}^{\mathrm{dn}}(s) \mathcal{H}_{12}^0$. 
Specifically, the bandwidth of the communication channel is designed based on the cut-off frequency $\omega_{\mathrm{H0}}$ and $\omega_D$ of the subsystems $\mathcal{H}_{12}^0$ and $\mathcal{G}^{\mathrm{dn}}(s)$. 
Since $\mathcal{H}^0_{12}(s)$ is a first-order lag system,  $\omega_{\mathrm{H0}}=\sqrt{3}k$ rad/s holds. 
Moreover, $\omega_D = 4.14\mu/L^2$ as shown in Table \ref{tab:cutoff}.
To satisfy the design requirement (a) that the control bandwidth must be $\omega \in \varpi$, we consider the conditions for $\omega_{\mathrm{H0}},\,\omega_D\geq 10^{-2}\, \mathrm{rad/s}$. 
Substituting the parameters for the other requirements (b) and (c) into $\omega_{\mathrm{H0}}$ and $\omega_D$, we obtain 
\begin{enumerate}[(i)]
\item $\mu\geq 24.2\,\si{\micro m^2\cdot s^{-1}}$
\item $k\geq 5.77\times 10^{-3}\,\si{s^{-1}}$.
\end{enumerate}

\smallskip
\par

\begin{figure*}
    \centering
    \includegraphics[width=0.85\linewidth]{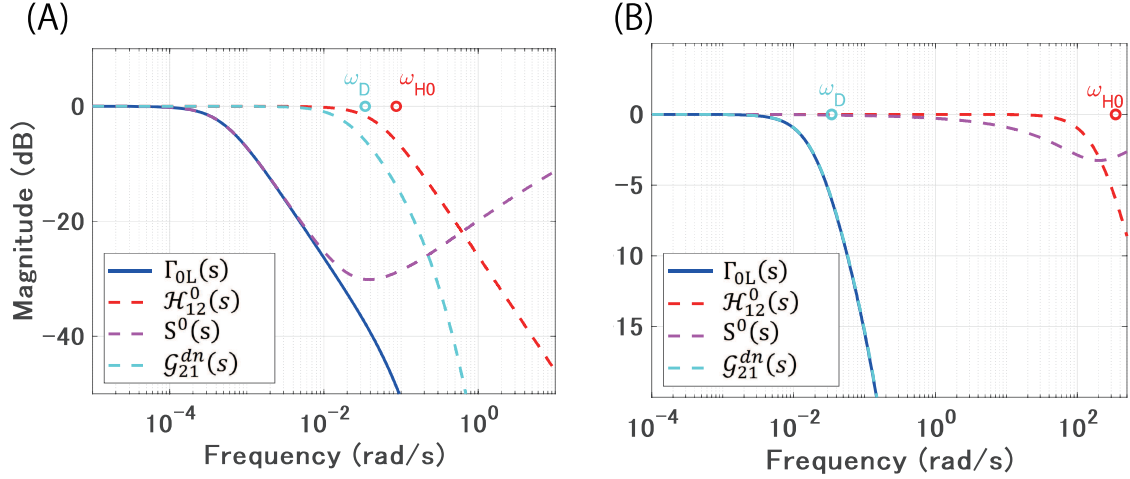}
    \caption{The gain diagram of the MC channel $\Gamma_{\mathrm{0L}}(j\omega)$, the boundary system $\mathcal{H}^0_{21}(j\omega)$, the self-interference system $S^0(j\omega)$, and the diffusion system $\mathcal{G}^{\mathrm{dn}}_{21}(j\omega)$ (A) for the membrane transport rate $k=5.0\times10^{-2}\,\si{s^{-1}}$ (B) for the membrane transport rate $k=200\,\si{s^{-1}}$.}
    \label{fig:wgain}
\end{figure*}

\smallskip
\par
The solid line in Fig. \ref{fig:wgain} (A) illustrates the gain of the MC channel $\Gamma_{\mathrm{0L}}(j\omega)$ for $k=5.0\times10^{-2}\,\si{s^{-1}}$, $\mu=83\,\si{\micro m^2\cdot  s^{-1}}$, and $L=100\,\si{\micro m}$, which satisfy the parameter conditions (i) and (ii). These parameter values are taken from those for quorum sensing \cite{Stewart2003,Pai2009}. 
As expected, the cut-off frequency of the boundary system $\mathcal{H}^0_{12}(j\omega)$ and the diffusion system $\mathcal{G}^{\mathrm{dn}}_{21}(j\omega)$ is designed to pass the signals for the desired control bandwidth $\omega \in \varpi$.  
However, Fig. \ref{fig:wgain} (A) shows that the gain of the MC channel $\Gamma_{\mathrm{0L}}(s)$ is attenuated between $10^{-3}\,\si{rad/s}$ and  $10^{-2}\,\si{rad/s}$ even though the parameters satisfy the conditions (i) and (ii).
This is because of the self-interference effect discussed in Section \ref{sec:boundarydiffusion}. 
Specifically, the gain of the self-interference system $S^0(j\omega)$, which has a large dip around $10^{-3}\leq\omega\leq10^{1}\, \si{rad/s}$ (see the dotted magenta line in Fig. \ref{fig:wgain}), restricts the bandwidth of the MC channel since the closed loop transfer function is $\Gamma_{\mathrm{0L}}(s) =
\mathcal{G}^{\mathrm{dn}}_{21}(s)S^0(s)\mathcal{H}^0_{12}(s)$. 
Thus, the effect of the self-interference system $S^0(j\omega)$ needs to be suppressed to design the MC channel $\Gamma_{\mathrm{0L}}(s)$ with a desired frequency response characteristics. 

\par
\smallskip
To suppress the effect of the self-interference system $S^0(s)$, we consider parameter conditions that prevent the gain of the self-interference system from being small. We suppose that signals do not significantly decrease if the gain of the self-interference system is $|S^0(j\omega)|\geq1/2$ for the control bandwidth $\omega \in \varpi$, which means $|\mathcal{H}^0_{11}(j\omega)\mathcal{G}^{\mathrm{dn}}_{11}(j\omega)|\leq1$ for $\omega \in \varpi$. However, since it is difficult to attenuate $|\mathcal{H}^0_{11}(j\omega)\mathcal{G}^{\mathrm{dn}}_{11}(j\omega)|$ only for $\omega \in \varpi$, we consider the entire frequency range. Using the approximation in Eq. (\ref{eq:tangain}) and the frequency characteristics of $\mathcal{H}^0_{11}(s)$ as a first-order lag system, the maximum gain of $\mathcal{H}^0_{11}(j\omega)\mathcal{G}^{\mathrm{dn}}_{11}(j\omega)$ can be approximated as
\begin{eqnarray}
\alpha(\omega)&:=&\max_{\omega}|\mathcal{H}^0_{11}(j\omega)\mathcal{G}^{\mathrm{dn}}_{11}(j\omega)| \nonumber\\
&\approx&
\left\{
\begin{array}{ll}
\frac{\sqrt{\mu}}{\sqrt{k}\Delta r} & (k \geq \frac{\mu}{L^2}) \\
\frac{L}{\Delta r} & (k < \frac{\mu}{L^2})
\end{array}
\right..\label{eq:hgmax}
\end{eqnarray}
If $\alpha(\omega)\leq1$, the gain of the self-interference system is $|S^0(j\omega)|\geq1/2$. Since the thickness $\Delta r$ of the small region at the boundary is $\Delta r<L$, the following condition is obtained to suppress the effect of the self-interference system $S^0(s)$.
\begin{enumerate}[(i)]
\setcounter{enumi}{2}
\item $k\geq \mu$
\end{enumerate}

\smallskip
\par
Fig. \ref{fig:wgain} (B) shows the gain characteristics of the MC channel for the redesigned membrane transport rate $k=200\,\si{s^{-1}}$, which satisfies both of the conditions (ii) and (iii). The value of the membrane transport rate $k$ is taken from that for ion channels \cite{Wang2019}. 
The gain of the redesigned MC channel $\Gamma_{\mathrm{0L}}(s)$ in Fig. \ref{fig:wgain} (B) becomes larger than $-6$ dB for the control bandwidth $\omega \in \varpi$. 
In particular, the loss of the gain in the self-interference system $S^0(s)$ that restricted the bandwidth of the MC channel is now shifted to the higher frequency range by suppressing the effect of the feedback interaction between the diffusion system and the boundary system (see the dotted magenta line in Fig. \ref{fig:wgain} (B)). 
Thus, the design specifications (a)--(c) are satisfied for the redesigned MC channel $\Gamma_{\mathrm{0L}}(s)$.

\begin{figure}
    \centering
    \includegraphics[width=0.99\linewidth]{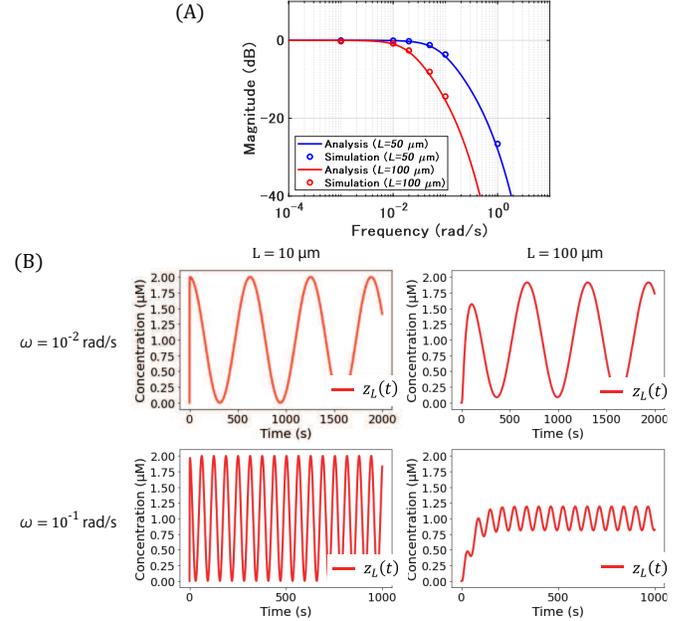}
    \caption{(A) The gain diagram of the MC channel $\Gamma_{\mathrm{0L}}(j\omega)$ for different communication distance $L=50\,\si{\micro m}$ and $L=100\,\si{\micro m}$. (B) Cosine wave inputs $c_0(t)$ and outputs $z_L(t)$ for $\omega=10^{-2}\,\si{rad/s}$ and $\omega=10^{-1}\,\si{rad/s}$ in time domain.}
    \label{fig:bode}
\end{figure}

\par
\smallskip
Finally, we verify that the analysis result of the frequency response characteristics agrees with the simulation in the time domain.  For this purpose, we compute the solution of the ODE model of the membrane kinetics (\ref{eq:transmembrane}) and the diffusion equation (\ref{eq:dif}) using finite difference methods implemented in Python. Fig. \ref{fig:bode} (A) shows the gain characteristics of the MC channels for different communication distances $L=50\,\si{\micro m}$ and $L=100\,\si{\micro m}$ with the same parameter values as in Fig. \ref{fig:wgain} (B). Since these parameter values satisfy the conditions (i) -- (iii), the gain of $\Gamma_{\mathrm{0L}}(j\omega)$ exceeds $-6$ dB for the control bandwidth $\omega \in \varpi$ for both of the distances, implying that the signal can be transmitted via the designed MC channel.
Fig. \ref{fig:bode} (B) shows the time series data of the output signal $z_L(t)$ at the molecular robot on the right for the cosine wave input with the frequency $\omega=10^{-2}\,\si{rad/s}$ and $\omega=10^{-1}\,\si{rad/s}$. When $\omega=10^{-2}\,\si{rad/s}$, the output signal $z_L(t)$ has approximately the same magnitude as the input signal $c_0(t)$ for the both distances. On the other hand, when $\omega=10^{-1}\,\si{rad/s}$, the magnitude of $z_L(t)$ is less than 1/4 of the magnitude of the input $c_0(t)$ for $L=100\,\si{\micro m}$ since the input frequency exceeds the pre-specified bandwidth $\omega \in \varpi$.
The gains calculated by the numerical simulations are plotted as the circular dots in Fig. \ref{fig:bode} (A). The simulated values agree with the theoretically predicted values, validating the frequency response analysis performed in this section. 
These results imply that the control-theoretic framework in Fig. \ref{fig:blockabstract} is a powerful tool for engineering MC channels in a system theoretic approach.

\subsection{Analysis of the entire MC channel $\mathcal{M}_{0L}(s)$ with ligand-receptor systems}
\label{sec:num-reception}

Finally, we consider the frequency response of the entire MC channel $\mathcal{M}_{0L}(s)$, {\it i.e.,} the transfer function from $c_0(s)$ to $y_L(s)$. 
The transfer function $\mathcal{M}_{0L}(s)$ involves the boundary system $\mathcal{H}_L(s)$, which is the  receptor-mediated signal transfer into the molecular robot modeled by Eq. (\ref{eq:receptor}), in addition to the MC channel $\Gamma_{0L}(s)$ in Fig. \ref{fig:blockabstract}. 
Thus, we need an additional analysis to make sure that the entire MC channel satisfies the pre-specified conditions (a)--(c) in Section \ref{sec:num-design}. In what follows, we show that, under a certain condition, the frequency response of the entire MC channel $\mathcal{M}_{0L}(s)$ can be designed by independently shaping the gain characteristics of the two transfer functions $\Gamma_{0L}(s)$ and $\mathcal{H}^L_{21}(s)$. 

\par
\smallskip
As shown in Fig. \ref{fig:blockabstract}, the input-output relation of $z_L(s)$ and $y_L(s)$ is characterized by the transfer function $\mathcal{H}^L_{21}(s)$ as $y_L(s)=\mathcal{H}^L_{21}(s)z_L(s)$. 
The input signal to the boundary system, $z_L(s)$, is subject to the self-interference effect of the boundary system $\mathcal{H}_L(s)$ since the input from the boundary to the diffusion region, $v_L(s)$, feeds back to $z_L(s)$. 
Thus, the transfer function from $c_0(s)$ to $z_L(s)$ involves not only $\Gamma_{0L}(s)$ but also the self-interference effect due to the reflection of the signal at the boundary $r=L$, which complicates the frequency response analysis of $\mathcal{M}_{0L}(s)$ in general.

\par
\smallskip
To closely look at the situation, we compute the transfer functions from the input $z_L(s)$ to the output $v_L(s) =  \partial c(s,L)/\partial r$ and obtain
\begin{eqnarray}
v_L(s) = \mathcal{H}^L_{11}(s)z_L(s) = \frac{Rk_{\mathrm{on}}}{\mu}\frac{s}{s+k_{\mathrm{off}}}z_L(s).\label{eq:vL}
\end{eqnarray}
A key observation here is that the transfer function $\mathcal{H}^L_{11}(s)$ is a high-pass filter with a cut-off frequency $\omega_{\mathrm{HL}} = k_{\mathrm{off}}/\sqrt{3}$, while $\Gamma_{0L}(s)$ is a low-pass filter as shown in Figs. \ref{fig:wgain} and \ref{fig:bode}. 
In other words, the frequency component less than $\omega_{\mathrm{HL}}$ is effectively eliminated when the signal is reflected at the boundary while the signal arriving at the boundary contains only the low frequency signal. 
Thus, $v_L(s)\approx 0$ if the gain for the product of the boundary system $\mathcal{H}^L_{11}(s)$ and the MC channel $\Gamma_{0L}(s)$ satisfies $|\mathcal{H}^L_{11}(j\omega)\Gamma_{0L}(j\omega)|\approx 0$. 
In this case, the transfer function from the sender to the receiver is calculated simply by the product of $\Gamma_{0L}$ and $\mathcal{H}^L_{21}(s)$, {\it i.e.,} $y_L(s)=\mathcal{H}^L_{21}(s)\Gamma_{0L}(s)c_0(s)$. 
This means that the gain of the entire MC channel $\mathcal{M}_{0L}(s)$ can be independently characterized by those of the two transfer functions, $|\mathcal{H}^L_{21}(s)|$ and $|\Gamma_{0L}(s)|$. 

\par
\smallskip
More specifically, let $\omega_M$ be defined as the dominant cut-off frequency of $\Gamma_{0L}(s)$, {\it i.e,} $\omega_{M} := \min(\omega_D, \omega_{\mathrm{H0}})$. Then, $|\mathcal{H}^L_{11}(j\omega)\Gamma_{0L}(j\omega)|\approx 0$ holds when $k_{\mathrm{off}}\gg\sqrt{3}\omega_M$. Thus, under this condition, the frequency response of the entire MC channel $\mathcal{M}_{0L}(s)$ can be analyzed independently by the MC channel $\Gamma_{0L}(s)$ and the ligand-receptor system $\mathcal{H}^L(s)$. 

\smallskip
\par
\begin{figure}
    \centering
    \includegraphics[width=0.99\linewidth]{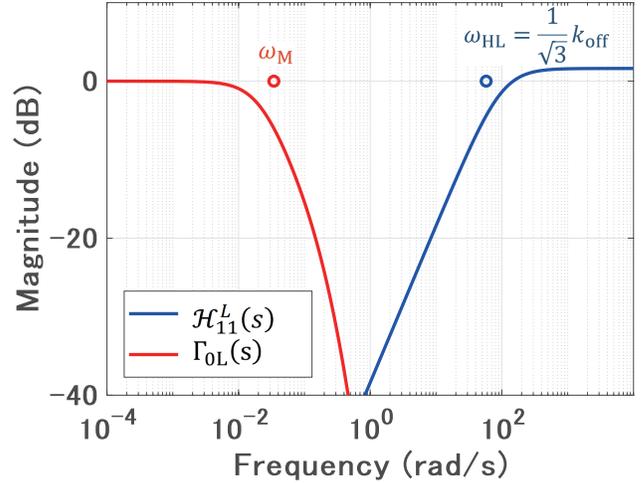}
    \caption{The gain diagram of the boundary system $\mathcal{H}^L_{11}(j\omega)$ and the MC channel $\Gamma_{\mathrm{0L}}(j\omega)$ with the cut-off frequencies.}
    \label{fig:yL}
\end{figure}

For the parameters $R=10^3$, $k_{\mathrm{on}}=10^{-1}\,\si{\micro M\cdot s^{-1}}$, and $k_{\mathrm{off}}=100\,\si{s^{-1}}$ \cite{Model1995,Andrews2009} in addition to the parameters of the MC channel $\Gamma_{0L}(s)$ used in Fig. \ref{fig:wgain} (B), we plot the gain diagram of the boundary system $\mathcal{H}^L_{11}(s)$ and the MC channel $\Gamma_{0L}(s)$ in Fig. \ref{fig:yL}. 
Since $k_{\mathrm{off}}=100\,\si{s^{-1}}$ and $\omega_{M} = \omega_D = 3.44\times 10^{-2}\,\si{rad\cdot s^{-1}}$, $k_{\mathrm{off}}\gg\sqrt{3}\omega_M$ holds, and $|\mathcal{H}^L_{11}(j\omega)\Gamma_{0L}(j\omega)| \approx 0$ for all frequencies. Therefore, the entire MC channel $\mathcal{M}_{0L}(s)$ can be designed based on the independent gain characteristics of $|\mathcal{H}^L_{21}(j\omega)|$ and $|\Gamma_{0L}(j\omega)|$, which means that the design specifications (a)--(c) are satisfied by designing them based on the conditions (i)--(iii) in Section \ref{sec:num-design} and the cut-off frequency of $|\mathcal{H}^L_{21}(j\omega)|$.

\section{Conclusion}
\label{sec:conclusion}

In this paper, we have proposed a control-theoretic modeling framework for bidirectional MC systems and performed the frequency response analysis to obtain a general design guideline of the MC channels and the boundary systems. Specifically, we have first derived the open loop characteristics of the diffusion system to design desired MC channels that can transfer signals with pre-specified control bandwidth. We have then shown the design guideline of the boundary system $\mathcal{H}^{*}(s)$ to suppress the undesirable effect of the self-interference. Finally, we have analyzed the frequency response characteristics of the specific MC channel to obtain the parameter conditions guaranteeing the performance of the system to satisfy the predefined specifications. 

\smallskip
\par
Using the proposed control-theoretic modeling framework, the stability, robustness and performance of the bidirectional MC systems can be analyzed based on the characteristics of each subsystem using methods in control theory. We expect that the proposed framework will be the foundation for the future development of feedback control theory for MC systems toward engineering applications.

\appendices

\section{Derivation of Theorem 1}
\label{sec:dtf}

We show the derivation of Theorem 1. By taking Laplace transform of Eq. (\ref{eq:dif}) for time $t$ and using the initial condition in Eq. (\ref{eq:initial}), we obtain
\begin{eqnarray}
sc(s,r) = \mu \frac{\partial^2 c(s,r)}{\partial r^2}.
\label{eq:tlaplace}
\end{eqnarray}
Similarly, the Laplace transform of Eq. (\ref{eq:tlaplace}) for the spatial variable $r$ leads to
\begin{eqnarray}
\frac{s}{\mu}c(s,p) = p^2 c(s,p)-c(s,0) - \frac{\partial c(s,0)}{\partial r},
\label{eq:rlaplace}
\end{eqnarray}
where $p$ is the frequency variables for space. Eq. (\ref{eq:rlaplace}) can be decomposed into the partial fractions
\begin{eqnarray}
c(s,p) &=& \frac{1}{2}\left(\frac{1}{p+\sqrt{\frac{s}{\mu}}}+\frac{1}{p-\sqrt{\frac{s}{\mu}}}\right)c(s,0) \label{eq:pfd}\\
&&+ \frac{1}{2\sqrt{\frac{s}{\mu}}}\left(\frac{1}{p-\sqrt{\frac{s}{\mu}}}-\frac{1}{p+\sqrt{\frac{s}{\mu}}}\right)\frac{\partial c(s,0)}{\partial r}.\nonumber
\end{eqnarray}
Taking the inverse Laplace transformation for space, we have the concentration $c(s,r)$ as
\begin{eqnarray}
c(s,r) &=& \frac{1}{2}\left(e^{-\sqrt{\frac{s}{\mu}}r}+e^{\sqrt{\frac{s}{\mu}}r}\right)c(s,0) \label{eq:pfd2}\\
&&+ \frac{1}{2\sqrt{\frac{s}{\mu}}}\left(e^{\sqrt{\frac{s}{\mu}}r}-e^{-\sqrt{\frac{s}{\mu}}r}\right)\frac{\partial c(s,0)}{\partial r}\nonumber\\
&=& \frac{1}{2}\left(g(s,r)+g(s,-r)\right)c(s,0) \nonumber\\
&&+ \frac{1}{2\sqrt{\frac{s}{\mu}}}\left(g(s,-r)-g(s,r)\right)\frac{\partial c(s,0)}{\partial r},\nonumber
\end{eqnarray}
and the concentration gradient $\partial c(s,r)/\partial r$ as
\begin{eqnarray}
\frac{\partial c(s,r)}{\partial r} &=& \frac{\sqrt{\frac{s}{\mu}}}{2}\left(e^{\sqrt{\frac{s}{\mu}}r} - e^{-\sqrt{\frac{s}{\mu}}r}\right)c(s,0) \label{eq:pfd3}\\
&&+ \frac{1}{2}\left(e^{\sqrt{\frac{s}{\mu}}r}+e^{-\sqrt{\frac{s}{\mu}}r}\right)\frac{\partial c(s,0)}{\partial r}\nonumber\\
&=& \frac{\sqrt{\frac{s}{\mu}}}{2}\left(g(s,-r) - g(s,r)\right)c(s,0) \nonumber\\
&&+ \frac{1}{2}\left(g(s,-r)+g(s,r)\right)\frac{\partial c(s,0)}{\partial r}.\nonumber
\end{eqnarray}
For the combination of the inputs $v_0(s)=c(s,0)$ and $v_L(s)=\partial c(s,L)/\partial r$, we transform Eq. (\ref{eq:pfd2}) to 
\begin{eqnarray}
c(s,r) &=& \frac{g(s,r) + g(s,L-r)g(s,L)}{1+g(s,L)g(s,L)}c(s,0) \label{eq:DNcsr}\\
&&+ \sqrt{\frac{\mu}{s}}\frac{- g(s,r)g(s,L) + g(s,L-r)}{1+g(s,L)g(s,L)}\frac{\partial c(s,L)}{\partial r}\nonumber
\end{eqnarray}
by substituting
\begin{eqnarray}
\frac{\partial c(s,0)}{\partial r} &=& - \sqrt{\frac{s}{\mu}}\frac{g(s,-L) - g(s,L)}{g(s,-L)+g(s,L)}c(s,0) \label{eq:pfd3-22}\\
&& + 2\frac{1}{g(s,-L)+g(s,L)}\frac{\partial c(s,L)}{\partial r}\nonumber
\end{eqnarray}
into Eq. (\ref{eq:pfd2}), where Eq. (\ref{eq:pfd3-22}) is obtained by substituting $r=L$ into Eq. (\ref{eq:pfd2}). Substituting $r=\ell$ into Eq. (\ref{eq:DNcsr}), we obtain
\begin{eqnarray}
c(s,\ell) &=& \frac{g(s,\ell) + g(s,L-\ell)g(s,L)}{1+g(s,L)g(s,L)}c(s,0) \label{eq:DNcsell}\\
&&+ \sqrt{\frac{\mu}{s}}\frac{- g(s,\ell)g(s,L) + g(s,L-\ell)}{1+g(s,L)g(s,L)}\frac{\partial c(s,L)}{\partial r}\nonumber\\
&=& G_{r,0}(s)|_{r=\ell}c(s,0) + \int^{\ell}_{0}{G_{r,L}(s)dr}\frac{\partial c(s,L)}{\partial r},\nonumber
\end{eqnarray}
where $G_{r,*}(s)$ is the transfer function from $c(s,*)$ to $c(s,r)$, which is same as the transfer function from $\partial c(s,*)/\partial r$ to $\partial c(s,r)/\partial r$. The reflection coefficient is $K=1$ because the boundary $\partial\Omega$ is free end relative to the input $v_*(s)$ for this combination of the inputs. The second term of Eq. (\ref{eq:DNcsell}) has the integral of $G_{r,L}(s)$ by $r$ since the output is the concentration while the input is the concentration gradient. On the other hand, the concentration gradient $\partial c(s,\ell)/\partial r$ is obtained by substituting Eq. (\ref{eq:pfd3-22}) into Eq. (\ref{eq:pfd3}) and $r=\ell$, that is
\begin{eqnarray}
\frac{\partial c(s,\ell)}{\partial r} &=& -\sqrt{\frac{s}{\mu}}\frac{-g(s,\ell) + g(s,L-\ell)g(s,L)}{1+g(s,L)g(s,L)}c(s,0) \nonumber\\
&&+ \frac{g(s,\ell)g(s,L) + g(s,L-\ell)}{1+g(s,L)g(s,L)}\frac{\partial c(s,L)}{\partial r}\label{eq:DNpcsell}\\
&=& \frac{\partial}{\partial r}G_{r,0}(s)\biggr|_{r=\ell}c(s,0) + G_{r,L}(s)|_{r=\ell}\frac{\partial c(s,L)}{\partial r}.\nonumber
\end{eqnarray}
The first term of Eq. (\ref{eq:DNpcsell}) has the derivative of $G_{r,0}(s)$ by $r$ since the output is the concentration gradient while the input is the concentration. Each term of Eqs. (\ref{eq:DNcsell}) and (\ref{eq:DNpcsell}) can be generally expressed by Eq. (\ref{eq:zr}), which can be applied to the concentration $c(s,\ell)$ and the concentration gradient $\partial c(s,\ell)/\partial r$ for the other input combinations in the same way.

\section{Proof of Proposition 1}

We prove that the gain of the diffusion transfer functions $\mathcal{G}_{21}^{\mathrm{dn}}(j\omega)$, $\mathcal{G}_{21}^{\mathrm{nd}}(j\omega)$, $\mathcal{G}_{21}^{\mathrm{dd}}(j\omega)$, and $\mathcal{G}_{21}^{\mathrm{nn}}(j\omega)$ monotonically decrease for $\omega\geq0$.

\subsection{The gain of the diffusion transfer functions $\mathcal{G}_{21}^{\mathrm{dn}}(j\omega)$ and $\mathcal{G}_{21}^{\mathrm{nd}}(j\omega)$}

We consider the gain of the diffusion transfer function 
\begin{eqnarray}
\left|\mathcal{G}_{21}^{\mathrm{dn}}(j\omega)\right| = \left|\mathcal{G}_{21}^{\mathrm{nd}}(j\omega)\right| = \frac{2}{\left(e^{\tilde{\omega}}+e^{-\tilde{\omega}}+2\cos{(\tilde{\omega})}\right)^{\frac{1}{2}}},\label{eq:gainnd}
\end{eqnarray}
where $\tilde{\omega} = \sqrt{2}L\sqrt{\omega}/\sqrt{\mu}$ for $\omega\geq0$. The derivative of the gain of $\mathcal{G}_{21}^{\mathrm{dn}}(j\omega)$ is
\begin{eqnarray}
\frac{d\left|\mathcal{G}_{21}^{\mathrm{dn}}(\tilde{\omega})\right|}{d\omega} &=& \frac{d\left|\mathcal{G}_{21}^{\mathrm{dn}}(\tilde{\omega})\right|}{d\tilde{\omega}} \frac{d\tilde{\omega}}{d\omega}\label{eq:derdn}\\
&=& -\frac{f_-(\tilde{\omega})}{\left(e^{\tilde{\omega}} + e^{-\tilde{\omega}} + 2\cos{(\tilde{\omega})} \right)^{\frac{3}{2}}} \frac{d\tilde{\omega}}{d\omega},\nonumber
\end{eqnarray}
where 
\begin{eqnarray}
f_-(\tilde{\omega}) = e^{\tilde{\omega}}-e^{-\tilde{\omega}}-2\sin{(\tilde{\omega})},\label{eq:f-}
\end{eqnarray}
and $d\tilde{\omega}/d\omega\geq0$. Since
\begin{eqnarray}
e^{\tilde{\omega}}+e^{-\tilde{\omega}}+2\cos{(\tilde{\omega})}\geq0 \label{eq:minuscos}
\end{eqnarray}
by the inequality of arithmetic and geometric means, and $f_-(\tilde{\omega})\geq0$ for $\omega\geq0$ (see Appendix \ref{sec:Anon}), Eq. (\ref{eq:derdn}) is non-positive for $\omega\geq0$. Therefore, the gain of $\mathcal{G}_{21}^{\mathrm{dn}}(j\omega)$ monotonically decreases for $\omega\geq0$.

\subsection{The gain of the diffusion transfer function $\mathcal{G}_{21}^{\mathrm{dd}}(j\omega)$}

We consider the gain of the diffusion transfer function 
\begin{eqnarray}
\left|\mathcal{G}_{21}^{\mathrm{dd}}(j\omega)\right| = \frac{\sqrt{2}\tilde{\omega}}{L\left(e^{\tilde{\omega}}+e^{-\tilde{\omega}}-2\cos{(\tilde{\omega})}\right)^{\frac{1}{2}}}\label{eq:gaindd}
\end{eqnarray}
for $\omega\geq0$. The derivative of the gain of $\mathcal{G}_{21}^{\mathrm{dd}}(j\omega)$ is
\begin{eqnarray}
\frac{d\left|\mathcal{G}_{21}^{\mathrm{dd}}(\tilde{\omega})\right|}{d\omega} &=& \frac{d\left|\mathcal{G}_{21}^{\mathrm{dd}}(\tilde{\omega})\right|}{d\tilde{\omega}} \frac{d\tilde{\omega}}{d\omega}\label{eq:derdd}\\
&=& \frac{h(\tilde{\omega})}{L\left(e^{\tilde{\omega}}+e^{-\tilde{\omega}}-2\cos{(\tilde{\omega})}\right)^{\frac{3}{2}}} \frac{d\tilde{\omega}}{d\omega},\nonumber
\end{eqnarray}
where
\begin{eqnarray}
h(\tilde{\omega}) &=& 2\left(e^{\tilde{\omega}}+e^{-\tilde{\omega}}-2\cos{(\tilde{\omega})}\right)\\\nonumber
&& - \tilde{\omega}\left(e^{\tilde{\omega}}-e^{-\tilde{\omega}}+2\sin{(\tilde{\omega})}\right).
\end{eqnarray}
Since
\begin{eqnarray}
e^{\tilde{\omega}}+e^{-\tilde{\omega}}-2\cos{(\tilde{\omega})}\geq0
\end{eqnarray}
for $\omega\geq0$ by the inequality of arithmetic and geometric means, Eq. (\ref{eq:derdd}) is non-positive for $\omega\geq0$ if $h(\tilde{\omega})\leq0$. In what follow, we take the first and the second derivative of $h(\tilde{\omega})$ as
\begin{eqnarray}
\frac{dh(\tilde{\omega})}{d\tilde{\omega}} &=& e^{\tilde{\omega}}(-\tilde{\omega}+1) + e^{-\tilde{\omega}}(-\tilde{\omega}-1)  \nonumber\\
&& +2\sin{(\tilde{\omega})} - 2\tilde{\omega}\cos{(\tilde{\omega})},\\
\frac{d^2h(\tilde{\omega})}{d\tilde{\omega}^2} &=& -\tilde{\omega}f_-(\tilde{\omega})\label{eq:Hdd2d}.
\end{eqnarray}
Since $f_-(t)\geq0$ (see Appendix \ref{sec:Anon}), we obtain
\begin{eqnarray}
\frac{d^2h(\tilde{\omega})}{d\tilde{\omega}^2}\leq0.
\end{eqnarray}
Moreover, we have $dh(\tilde{\omega})/d\tilde{\omega}\leq0$ for $\omega\geq0$ because $d^2h(0)/d\tilde{\omega}^2=0$. Using $h(0)/d\tilde{\omega}=0$, we obtain $h(\tilde{\omega})\leq0$ for $\omega\geq0$, and Eq. (\ref{eq:derdd}) is non-positive. Therefore, the gain of the diffusion transfer function $|\mathcal{G}^{\mathrm{dd}}_{21}(j\omega)|$ monotonically decreases for $\omega\geq0$.

\subsection{The gain of the diffusion transfer function $\mathcal{G}_{21}^{\mathrm{nn}}(j\omega)$}

We consider the gain of the diffusion transfer function 
\begin{eqnarray}
\left|\mathcal{G}_{21}^{\mathrm{nn}}(j\omega)\right| = \frac{2\sqrt{2}L}{\tilde{\omega}\left(e^{\tilde{\omega}}+e^{-\tilde{\omega}}-2\cos{(\tilde{\omega})}\right)^{\frac{1}{2}}}\label{eq:prop}
\end{eqnarray}
for $\omega\geq0$. The derivative of the gain of $\mathcal{G}_{21}^{\mathrm{nn}}(j\omega)$ is
\begin{eqnarray}
\frac{d\left|\mathcal{G}_{21}^{\mathrm{nn}}(\tilde{\omega})\right|}{d\omega} &=& \frac{d\left|\mathcal{G}_{21}^{\mathrm{nn}}(\tilde{\omega})\right|}{d\tilde{\omega}} \frac{d\tilde{\omega}}{d\omega}\label{eq:dernn}\\
&=& -\frac{\sqrt{2}L}{\tilde{\omega}^2}\frac{2C_-(\tilde{\omega}) + \tilde{\omega}f_+(\tilde{\omega})}{C^{\frac{3}{2}}_-(\tilde{\omega})} \frac{d\tilde{\omega}}{d\omega},\nonumber
\end{eqnarray}
where 
\begin{eqnarray}
f_+(\tilde{\omega}) &=& e^{\tilde{\omega}} - e^{-\tilde{\omega}} + 2\sin{(\tilde{\omega})},\label{eq:f+} \\
C_- &=& e^{\tilde{\omega}} + e^{-\tilde{\omega}} - 2\cos{(\tilde{\omega})}.
\end{eqnarray}
Since $C_-(\tilde{\omega})\geq0$ by the inequality of arithmetic and geometric means, and $f_+(\tilde{\omega})\geq 0$ for $\omega\geq0$ (See Appendix \ref{sec:Anon}), Eq. (\ref{eq:dernn}) is non-positive for $\omega\geq0$. Therefore, the gain of $\mathcal{G}_{21}^{\mathrm{nn}}(j\omega)$ monotonically decreases for $\omega\geq0$.

\section{Non-negativity of Eq. (\ref{eq:f-}) and Eq. (\ref{eq:f+})}
\label{sec:Anon}

We consider the function
\begin{eqnarray}
f_{\pm}(\tilde{\omega}) &=& e^{\tilde{\omega}} - e^{-\tilde{\omega}} \pm 2\sin{(\tilde{\omega})},\label{eq:fpm}
\end{eqnarray}
and prove $f_{\pm}(\tilde{\omega})\geq0$ for $\tilde{\omega}\geq0$. We consider the derivative of $f_{\pm}(\tilde{\omega})$ as
\begin{eqnarray}
\frac{df_{\pm}(\tilde{\omega})}{d\tilde{\omega}} &=& e^{\tilde{\omega}} + e^{-\tilde{\omega}} \pm 2\cos{(\tilde{\omega})} .\label{eq:derfpm}
\end{eqnarray}
Using the inequality of arithmetic and geometric means, we obtain that Eq. (\ref{eq:derfpm}) is non-negative for $\tilde{\omega}\geq0$. Since $f_{\pm}(0)=0$, Eq. (\ref{eq:fpm}) is non-negative for $\tilde{\omega}\geq0$.


\vskip -2\baselineskip plus -1fil

\begin{IEEEbiography}[]{Taishi Kotsuka}
is a Ph.D. student in the Department of Applied Physics and Physico-Informatics, Keio University, where he received the B.S. degree and the M.S. degree in engineering in 2018 and 2021, respectively. He received the M.S. degree in science from Technical University Munich in 2021. His research interests include the applications of control theory to molecular communication systems. 
\end{IEEEbiography}

\vskip -2\baselineskip plus -1fil

\begin{IEEEbiography}[]{Yutaka Hori}
received the B.S degree in engineering, and the M.S. and Ph.D. degrees in information science and technology from the University of Tokyo in 2008, 2010 and 2013, respectively. He held a postdoctoral appointment at California Institute of Technology from 2013 to 2016. In 2016, he joined Keio University, where he is currently an associate professor. His research interests lie in feedback control theory and its applications to synthetic biomolecular systems. He is a recipient of Takeda Best Paper Award from SICE in 2015, and Best Paper Award at Asian Control Conference in 2011, and is a Finalist of Best Student Paper Award at IEEE Multi-Conference on Systems and Control in 2010. He has been serving as an associate editor of IEEE Transactions on Molecular, Biological and Multi-Scale Communications. He is a member of IEEE, SICE, and ISCIE.
\end{IEEEbiography}

\end{document}